# The Structure and Low-Energy Phonons of the Nonferroelectric Mixed Perovskite: $BaMg_{1/3}Ta_{2/3}O_3$.

A. Cervellino[1,2], S. N. Gvasaliya[2,3], B. Roessli[2], G. M. Rotaru[2], R. A. Cowley[4], S. G. Lushnikov[5], T. A. Shaplygina[5], and L. Bouchenoire[6,7]

[1] Swiss Light Source, Paul Scherrer Institute, CH-5232, Villigen, Switzerland.
[2] Laboratory for Neutron Scattering, PSI, CH-5232, Villigen, Switzerland.
[3] Laboratory for Solid State Physics, ETH Hönggerberg, 8093 Zürich, Switzerland
[4] Clarendon Laboratory Department of Physics, Oxford University, Parks Road, Oxford, OX1 3PU, UK
[5] Ioffe Physical Technical Institute, 194021, St Petersburg, Russia.
[6] European Synchrotron Radiation Facility, BP 220, F-38043 Grenoble CEDEX, France
[7] Department of Physics, University of Liverpool, Liverpool L69 7ZE, UK

## Abstract

The structure of $BaMg_{1/3}Ta_{2/3}O_3$ (BMT) has been studied using X-ray scattering. The phonons have been measured and the results are similar to those of other materials with the perovskite structure such as $PbMg_{1/3}Nb_{2/3}O_3$ (PMN). The acoustic and lowest energy optic branches were measured but it was not possible to measure the branches of higher energy, possibly this is because they largely consist of oxygen motions. High-resolution inelastic measurements also showed that the diffuse scattering was strictly elastic and not directly related to the phonon spectra. A diffuse scattering was observed in BMT near the (H±1/2, K±1/2, L±1/2) points in the Brillouin zone and this had a characteristic cube shape. This arises from ordering of the B-site ions in BMT. Additional experiments revealed a diffuse scattering in BMT similar in shape to Bragg reflections at wave-vectors of the form (H±1/3, K±1/3, L±1/3). Such reflections were also observed by Lufaso [Chem. Matt. **16** (2004) 2148] from powders and suggest that this structure of BMT consists of 4 differently oriented domains of a trigonal structure and results from a different ordering of the B-site ions from that responsible for the scattering at the (H±1/2, K±1/2, L±1/2) points. The results lead us to suggest that for BMT single crystals the bulk has the properties of a cubic perovskite, whereas the surface may have quite different structure from that of the bulk. This difference resembles the behaviour of cubic relaxors like PMN and PMN doped by $PbTiO_3$, where significant surface effects have been reported.

Corresponding Author: S. N. Gvasaliya (sgvasali@phys.ethz.ch).

## 1. Introduction

BaMg$_{1/3}$Ta$_{2/3}$O$_3$ (BMT) is a cubic material and has the perovskite structure, ABO$_3$, with the Mg and Ta ions distributed over the B sites. In this respect it is similar to the relaxor ferroelectrics like PbMg$_{1/3}$Nb$_{2/3}$O$_3$ (PMN) or PbMg$_{1/3}$Ta$_{2/3}$O$_3$ (PMT) which have disordered Mg/Nb or Mg/Ta on the B site [1, 2]. However, unlike relaxor ferroelectrics, BMT shows no evidence of ferroelectric behaviour and most of the properties of BMT change little as the temperature is varied. Furthermore, X-ray scattering measurements show that BMT has considerable diffuse scattering that is very different from that observed in PMN and related materials. The diffuse scattering in BMT depends significantly on the treatment of the sample. In some of the experiments the diffuse scattering from BMT consists of cubes of scattering centred on the points (H±1/2, K±1/2, L±1/2) in reciprocal space where we have labelled the points with the indices of the ideal cubic structure [3]. While in other experiments [4], the diffuse X-ray scattering is different and consists largely of sharp reflections that together with the strong Bragg reflections give rise to a rhombohedral structure. The structure then has additional Bragg reflections along the four [1, 1, 1] directions of the cubic perovskite structure with Bragg reflections at the cubic positions and at positions such as (H±1/3, K±1/3, L±1/3)$2\pi/a$, where $a$ is the cubic lattice parameter. In contrast to BMT, the diffuse scattering from cubic relaxors like PMN or PMT has two components: low energy quasi-elastic scattering centred largely at small wave-vectors transfers and strictly elastic scattering that tends to become larger at temperatures below the random-field (relaxor) transition [2].

Neutron scattering has been used to measure the intensities of the Bragg reflections for several relaxors. In particular we have examined the intensity of 632 Bragg reflections (128 of them are independent) of PMT at T = 300 K and T = 20 K using the D9 instrument at the ILL. The results were used to obtain the Patterson function for PMT and this showed that the Pb atoms are displaced away from their cubic positions as has been reported elsewhere [5]. The results are similar to those obtained for the Pb atoms in PMN, see Ref. [2] for a discussion. A difficulty with the BMT samples is that single crystals can only be obtained as small crystals with typical sizes of less than 1 mm$^3$. A consequence is that neutron scattering cannot be used to measure the inelastic scattering or the details of the diffuse scattering. We have therefore used a variety of X-ray scattering techniques to study BMT. Inelastic X-ray scattering enabled the measurement of the low energy parts of the phonon spectra. These measurements showed that the phonon spectra were qualitatively similar to the phonon spectra of other perovskites and that the phonon spectra had no anomalies at the wave-vectors corresponding to the maxima of the diffuse scattering. The diffuse scattering was measured and within the resolution of the instrument the elastic scattering did not show any broadening. These measurements are reported in more detail in section 2.

In octahedral environment the ionic radii for Mg and Ta differ only by ~5% and this ratio is very similar for Mg and Nb in PMN. As a result the different electrostatic charges are one of the dominant forces controlling the ordering on the B site of BMT. On the A site the charge on the Pb atoms for PMN is possibly less fixed at two electrons than the charge of the Ba atoms for BMT. We shall now discuss the different types of behaviour that can arise from the disorder on the B site. To preserve electrical neutrality one third of the B sites must be occupied by Mg ions and two thirds by Ta ions. There are four different types of structure that can arise associated with the B sites:

i) The B sites may separate into large regions enriched in either Mg or Ta ions. This should give rise to split or broadened Bragg reflections as a result of the slightly different radii of the Mg and Ta. This structure would be very unfavourable energetically due to the complete separation of the two differently charged atoms.

ii) The Mg and Ta ions are arranged homogeneously and randomly. This structure gives sharp Bragg peaks from the average perovskite structure and constant or slowly varying diffuse scattering from the disorder (a Laue-like scattering). This is in contrast to the observed diffuse scattering in BMT.

iii) Long range ordering of the Mg and Ta ions could occur according to the stoichiometry of BMT. This will give rise to additional Bragg peaks and so might be consistent with those X-ray measurements that suggest a rhombohedral structure.

iv) There could be a correlated distribution with the occupation of Mg and Ta being a well-defined function of the distance and the direction but without any long range periodicity. This type of structure is consistent with the structured diffuse scattering.

In section 3 we discuss the way the diffuse scattering was measured and then suggest that the reason for the different results is consistent with the structure being different at the surface from that of the bulk of the crystal. This type of behaviour has been found to occur in PMN and other relaxors [2] and is characteristic of these disordered oxides.

The shape of the diffuse scattering from the interior of the crystal has been measured and several different ways have been used to determine the structure responsible for this scattering. We shall briefly review the different approaches in section 4 but the full accounts are or will be published elsewhere [3, 5].

## 2. Inelastic X-ray Scattering

The inelastic X-ray scattering (IXS) experiments were performed on the ID28 beamline at the ESRF [6]. The instrument was operated with the incident energy of the X-rays is 17.794 keV providing a resolution of 3.0 meV full-width at half-maximum (FWHM). Direction and size of the momentum transfer were selected by an appropriate choice of the scattering angle and the crystal orientation in the horizontal scattering plane defined by the reciprocal space vectors [H,H,0] and [0,0,L]. The momentum resolution was set to ~0.25 nm$^{-1}$ x 0.75 nm$^{-1}$ in the horizontal and vertical plane, respectively. Further details on the spectrometer operation and the data treatment procedure are given in Ref. [6].

Typical results of constant wave-vector transfer experiments are shown in Figs. 1 and 2. Both show an intense elastic peak and a weaker inelastic peak with an energy of 18 meV, Fig. 1, and 15 meV, Fig. 2. The inelastic peaks arise from phonon scattering while the elastic scattering is from the diffuse scattering. The phonon shown in Fig.1 is expected to be longitudinally polarised. The low energy dispersion relations are shown in Fig. 3 and the results for the [0, 0, q] direction show dispersion curves for both transverse (TA) and longitudinal (LA) branches of the acoustic and lowest energy transverse (TO) and longitudinal (LO) optic modes. The energies of the modes are similar to the phonon energies in other oxide perovskites. A noticeable distinction of BMT from the ferroelectric perovskites is the nature of the lowest TO phonon branch near the Brillouin zone center. In a ferroelectric one expects a steep decrease of the soft phonon frequency as the wavevector approaches zero. As such behaviour is not observed in BMT, one may conclude that there is no soft optic phonon in BMT. Attempts were made to observe the optic modes of BMT at higher frequencies but these were unsuccessful. Possibly this is because these high energy modes are largely motions of the oxygen atoms and these are expected to have much lower intensity, due to the lower number of electrons, than the modes in which the heavy atoms are in motion. It is, however, of note that similar results were obtained for PMN by neutron scattering [2] and so the difficulties in identifying the scattering from high-frequency phonons could be characteristic of these disordered perovskites.

The IXS spectra Figs. 1 and 2 show that the width of the diffuse scattering is limited by the energy-resolution and also that its intensity is larger at the (q, q, q) zone boundary than along the [0, 0, q] direction as found in other measurements of the diffuse scattering from BMT.

### 3. Diffuse X-ray Scattering

To study the distribution of the diffuse scattering in the reciprocal space of BMT three different diffractometers were used: the Swiss-Norwegian Beam Lines [7] at the ESRF, and the XMAS [8] beam line at the ESRF and the PXI-X06SA [9] beam line at the Swiss Light Source (SLS). The instrument at the Swiss-Norwegian beam line (SNBL) was used with an incident X-ray energy of 17.45 keV and the scattered X-rays were recorded with a MAR345 image plate detector. The follow-up experiments were performed at XMAS and X06SA. The incident X-ray energy at the XMAS instrument was 10 keV and a Ge(333) analyzer together with a point detector was used to record the scattered X-rays. The instrument at the SLS had an incident X-ray energy of 8 keV and had a PILATUS 6M detector to record the scattered X-rays. For the area detector data the orientation matrix refinement and preliminary reciprocal space reconstructions were performed using the CrysAlis software package. For the final reciprocal space reconstructions we applied corrections for polarization and for solid angle conversion associated with the planar projection. Laue symmetry of the sample was applied to the reconstructed reciprocal space layers, thus improving the signal-to-noise ratio and removing the gaps between the individual detector elements in case of PILATUS 6M detector. The as grown samples were used for the experiments performed at the ESRF. Whereas prior to the experiment at the SLS a rod-like optically homogeneous sample with ~100 μm diameter was produced and etched with hot concentrated HCl in order to remove the (damaged) surface layer. SNBL and XMAS datasets were taken at room temperature, and in order to reduce the phonon contribution the experiment at the SLS was performed at 100 K.

The results obtained at the Swiss-Norwegian beam line and at the SLS were very similar to one another. The selected cuts of the reciprocal space are shown in Figs. 4 and 5. Figure 4 shows the scattering from regions of reciprocal space around the (2, 0, 0) and (2, 2, 0) Bragg reflections. For comparison, the scattering from the same regions is also shown in Fig. 4 for PMT. In the case of BMT, Fig. 4 shows that the scattering consists of dots from Bragg reflections broadened by the experimental resolution and by thermal diffuse scattering but unlike the scattering from PMT there is no evidence of the streaks found in PMT and that change in shape from one Bragg reflection to another.

The diffuse scattering from a larger region of reciprocal space is shown for the (H, K, 0.5) plane of BMT in Fig. 5. The diffuse scattering is approximately a regular set of squares centred on the positions with wave-vectors (H±1/2, K±1/2, L±1/2) in the Brillouin zone of the Pm3m structure. The results obtained with the XMAS instrument gave somewhat different results as shown in Fig. 6. Although there is an evidence for diffuse scattering occurring at the same wave-vectors as in the data taken at the SNBL and the X06SA beam line, the most intense features are peaks in the scattering that are much sharper and more similar to the Bragg reflections. The peaks are shown in Fig. 6 and are situated at wave vectors such as (H±0.33, K±0.33, L±0.33).

The X-ray results shown in Fig. 4 demonstrate that the diffuse scattering near the Bragg reflections is much weaker in BMT than for PMT. This implies that the displacements of Pb along the [1, 1, 1] directions identified in cubic relaxors, are absent in BMT. Consequently, the difference in the physical properties of BMT and PMN is due to the replacement of Pb by Ba which supresses the Pb driven nano-regions and hence changes the dielectric properties of BMT when compared with those of PMN or PMT.

The diffuse scattering shown in Fig. 5 shows a regular arrangement of cubes centred on the (H±1/2, K±1/2, L±1/2) positions. They have been explained in terms of the partial order on the B sites which shows that the electrostatic forces lead to short range order. The analysis of the structure can only be completed by using the technique of computer simulation, as described in references [3, 10]. We are unaware of other disordered perovskites that also show this type of structure. This is surprising as the charges are similar to those in other related materials and the electrostatic forces are the dominant interactions in them. Hence we suggest that the other B-site disordered cubic perovskites with Pb replaced by Ba might show similar diffuse scattering.

Another type of X-ray diffuse scattering was obtained using the XMAS instrument to study single crystals and consists of sharp peaks that at least resemble Bragg reflections at positions similar to (H±1/3, K±1/3, L±1/3) as shown in Fig. 6. These results are similar to those obtained when powdered samples of BMT are studied with X-rays scattering. Lufaso [11] studied the structure of BMT using powdered samples by X-ray and neutron scattering techniques and found that appropriately annealed BMT has the P-3m1 trigonal structure. This structure is formed from the cubic perovskite structure by ordering of the atoms on the B sites perpendicular to one of the (1, 1, 1) planes. Since there are 4 different [1, 1, 1] directions the samples probably consist of 4 different domains and these give rise to Bragg reflections in the cubic system at positions such as (H±1/3, K±1/3, L±1/3) as observed with the X-ray scattering obtained using the XMAS beam line.

A powder diffraction experiment was also performed with BMT and an incident X-ray energy of 17.522 keV at the SLS beam line MS-X04SA [12]. The BMT powder was not annealed and together with the intense Bragg reflections of the cubic structure, additional reflections were also observed at the positions (H±1/3, K±1/3, L±1/3). These reflections had a larger width than the cubic ones, corresponding to a correlation length of 20 nm. This observation is in-line with the results of Lufaso [11] in that our BMT powder that was not annealed shows only short-range rhombohedral order, possibly restricted to the grain surfaces. The results for powders and single crystal samples of BMT are clearly different. This is possibly because the powders have much larger surface areas and the rhombohedral phase only occurs at the surface of the rapidly quenched samples. This is similar to the behaviour of PMN where at low temperatures the ferroelectric structure only occurs near the surface when a sufficiently strong electric field is applied.

## 4. Conclusions

The structure of single crystals of BMT has been studied using X-ray scattering. Inelastic X-ray scattering has been performed to study the phonon spectra and the results for the acoustic and lowest energy optic modes are qualitatively similar to the phonons observed from conventional oxide perovskite materials.

No evidence was found for a zone centre soft optic mode. This is consistent with nearly temperature-independent and a low value of the dielectric constant of BMT. The experiments did not reveal scattering from the highest energy optic modes. Possibly this is because of the weak scattering from the oxygen atoms and the eigenvectors of high-frequency modes are likely to consist of oxygen vibrations. This observation for BMT is, however, similar to the neutron scattering results for the related relaxor PMN, where the high-frequency optic phonons could not be properly identified. The diffuse scattering from BMT was also measured and found to be strictly elastic within the experimental resolution of 3 meV.

The strong diffuse scattering that occurs close to the Bragg reflections for PMN and PMT was not observed for BMT. This shows that the polar nano-regions (regions where displacements of the Pb are large and correlated) are not formed and this is presumably the reason for the absence of relaxor properties in BMT. The diffuse scattering in BMT appears as cubes centred on the (H±1/2, K±1/2, L±1/2) positions of the perovskite structure. These peaks are believed to arise from partial ordering of the B site cations and the resulting structure and its thermodynamics has been discussed in detail [3, 10].

The results obtained from the lower energy instrument at the XMAS beam line are similar to those of Lufaso [11] who studied a powder of BMT. He suggests that the structure of powder is trigonal and has a unit cell that contains 3 formula units of the perovskite. This is different from the results described above for single crystals. We consider that this is because he used either an annealed powder of BMT or X-rays of low energy that are only able to penetrate the surface of the sample. In contrast, the single crystals of cubic BMT which we used in the experiments described above were synthesized by rapid quenching according to the procedure described in Ref. [13]. As discussed by Cervellino et al. [10] excessive entropy appears to be essential for stabilizing the cubic form of BMT. In addition, in most of the experiments with higher energy X-rays the surface had been very carefully prepared before the experiment. Both of these strongly influence the detection of surface-related effects. We therefore consider that the structure of single crystal BMT consists of a core, for which there is evidence of B site short range ordering from the observation of diffuse scattering around (H±1/2, K±1/2, L±1/2). The surface of the samples depends on the method of preparation but there is no doubt that it is possible that further B site ordering can occur and that for annealed fine powders the whole of powder grains can have a trigonal structure. In larger samples we suggest that the surface can also have this trigonal structure. Experiments on PMN have also shown that the surface of single crystals can be different from those of the bulk. In this case the distorted electric-field induced structure is rhombohedral with one formula unit in each unit cell. BMT is different because the distorted structure is related to the ordering of the B site and the structure is trigonal with 3 formula units in each unit cell.

## 5. Acknowledgements

We are grateful for the opportunity to carry out the measurements using the instruments: SNBL, ID28, and XMAS at the ESRF, and X04SA and X06SA at the SLS. We are indebted to Alexei Bosak and Dmitry Chernyshev for their help in the synchrotron experiments and for useful discussions. We are thankful to Oksana Zaharko for analysis and subsequent discussions on the results obtained at D9, to Maria Teresa Fernandez-Diaz for her help in the experiment at D9, and to Clemens Shultze-Briese (SLS, presently Dectris Ltd.) for his help in the experiment at X06SA.

9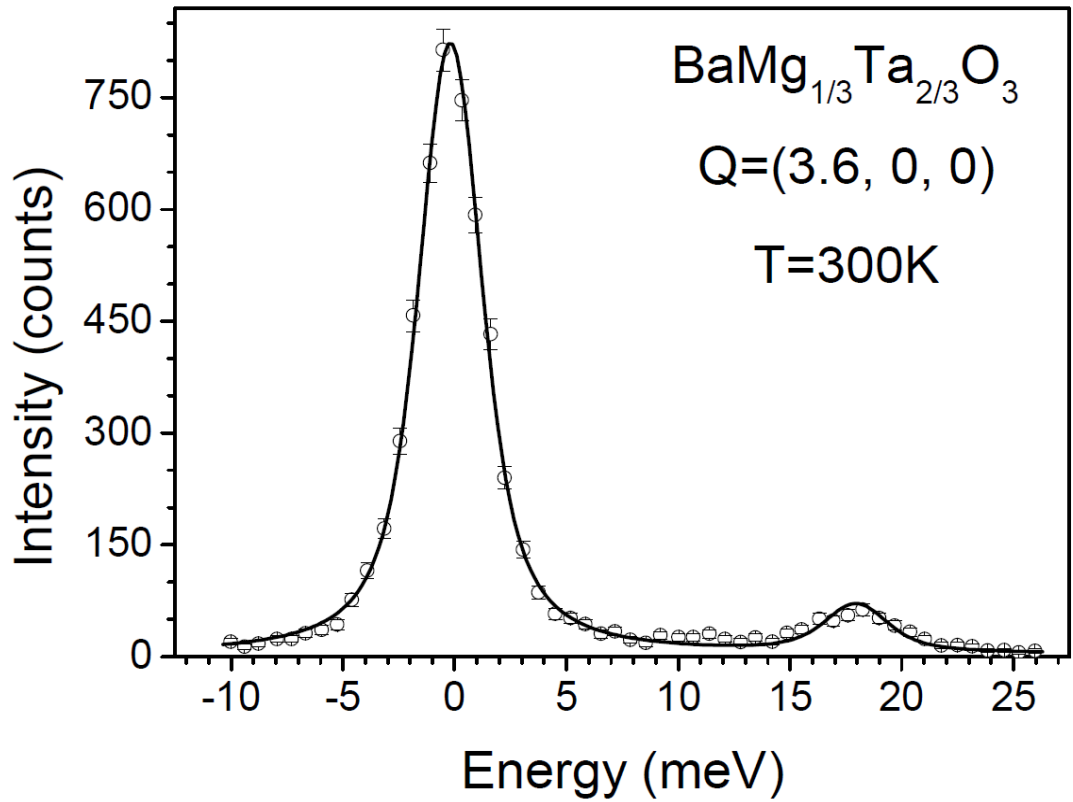

Figure 1. Constant-Q inelastic X-ray scattering scan from BMT with (3.6 0 0) momentum transfer. Solid line is a fit by two Lorentzians convoluted with the resolution function. The elastic peak has the energy-resolution width. The weaker peak is scattering from a phonon.

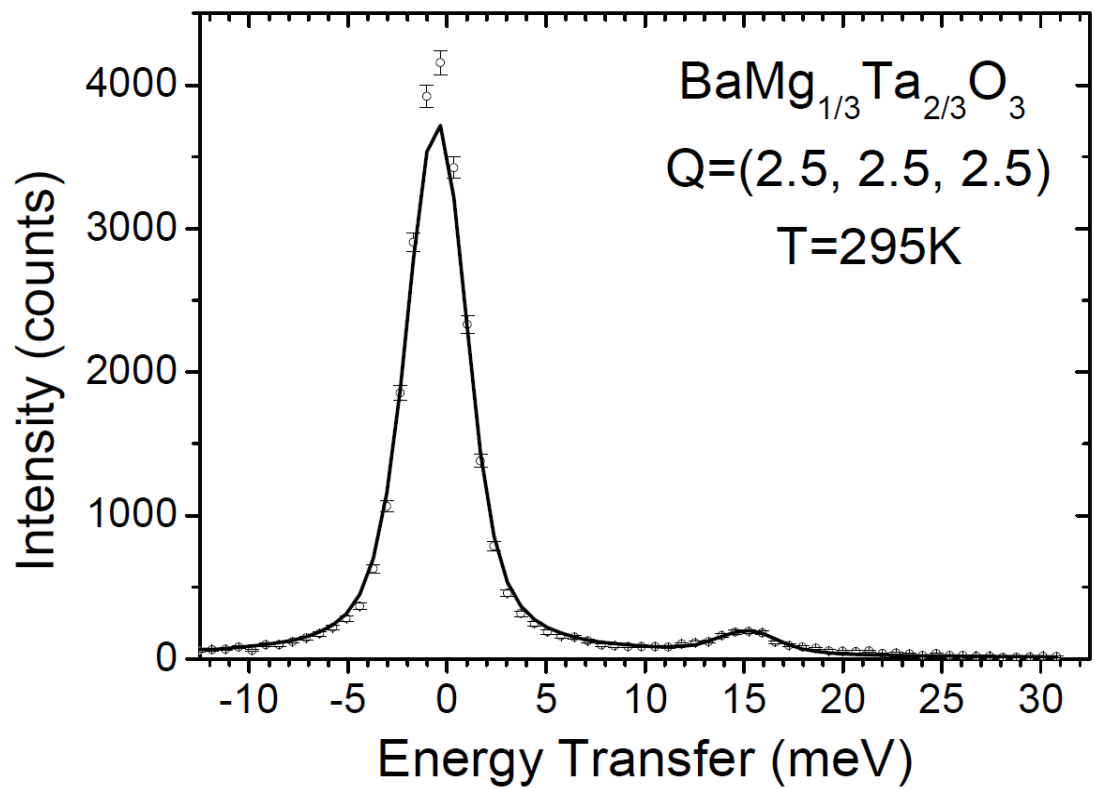

Figure 2. Constant-Q inelastic X-ray scattering scan from BMT with (2.5, 2.5, 2.5) momentum transfer. Solid line is a fit by two Lorentzians convoluted with the resolution function. The elastic peak has the energy-resolution width. The weaker peak is scattering from a phonon.
.

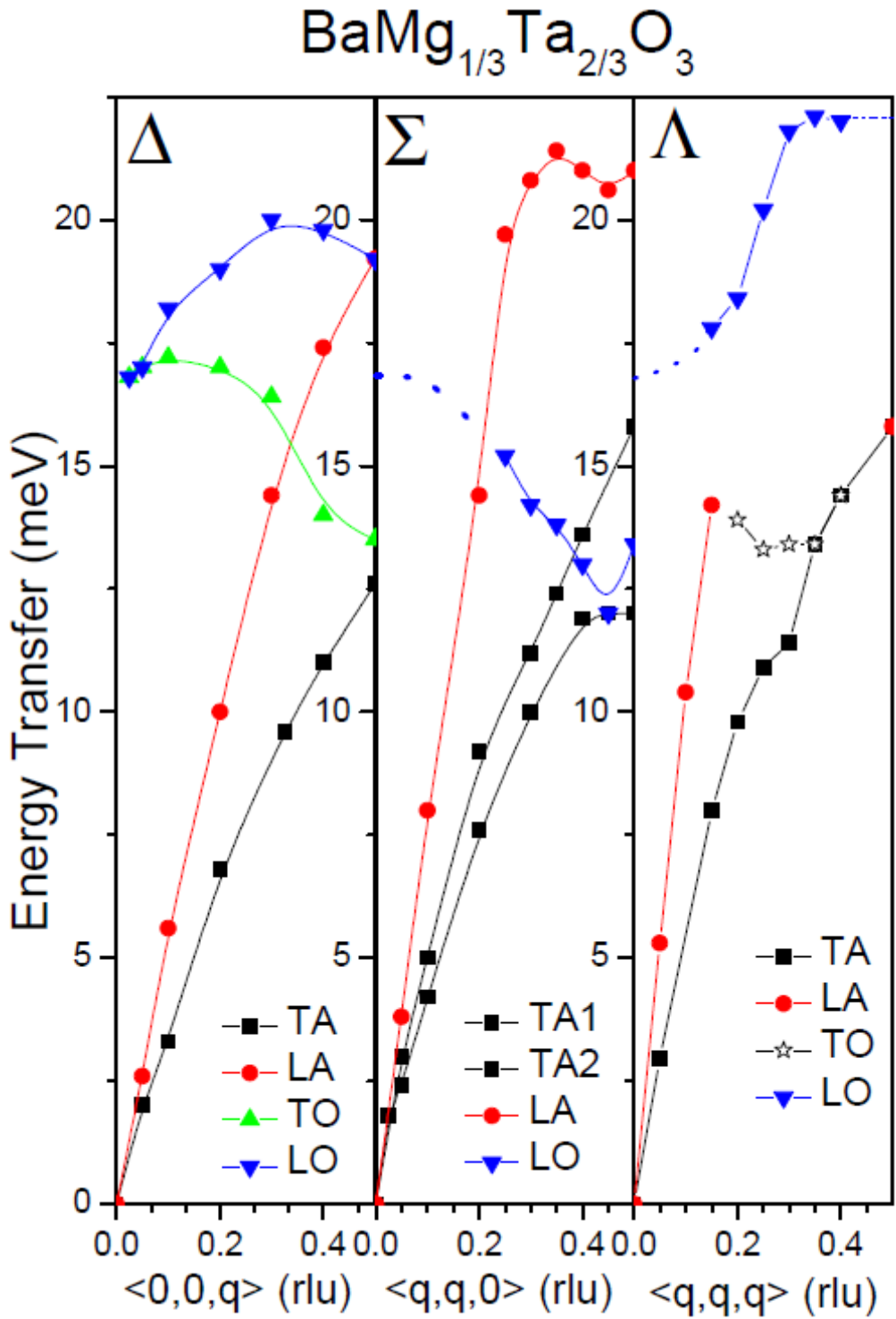

Figure 3. (Color online) Dispersion curves for the phonons in BMT and propagating along the 3 high symmetry directions <q, 0, 0>, <q, q, 0> and <q, q, q>. The stars on the right panel are probably due to the TO phonon. The dotted lines show extrapolation to the zone centers.

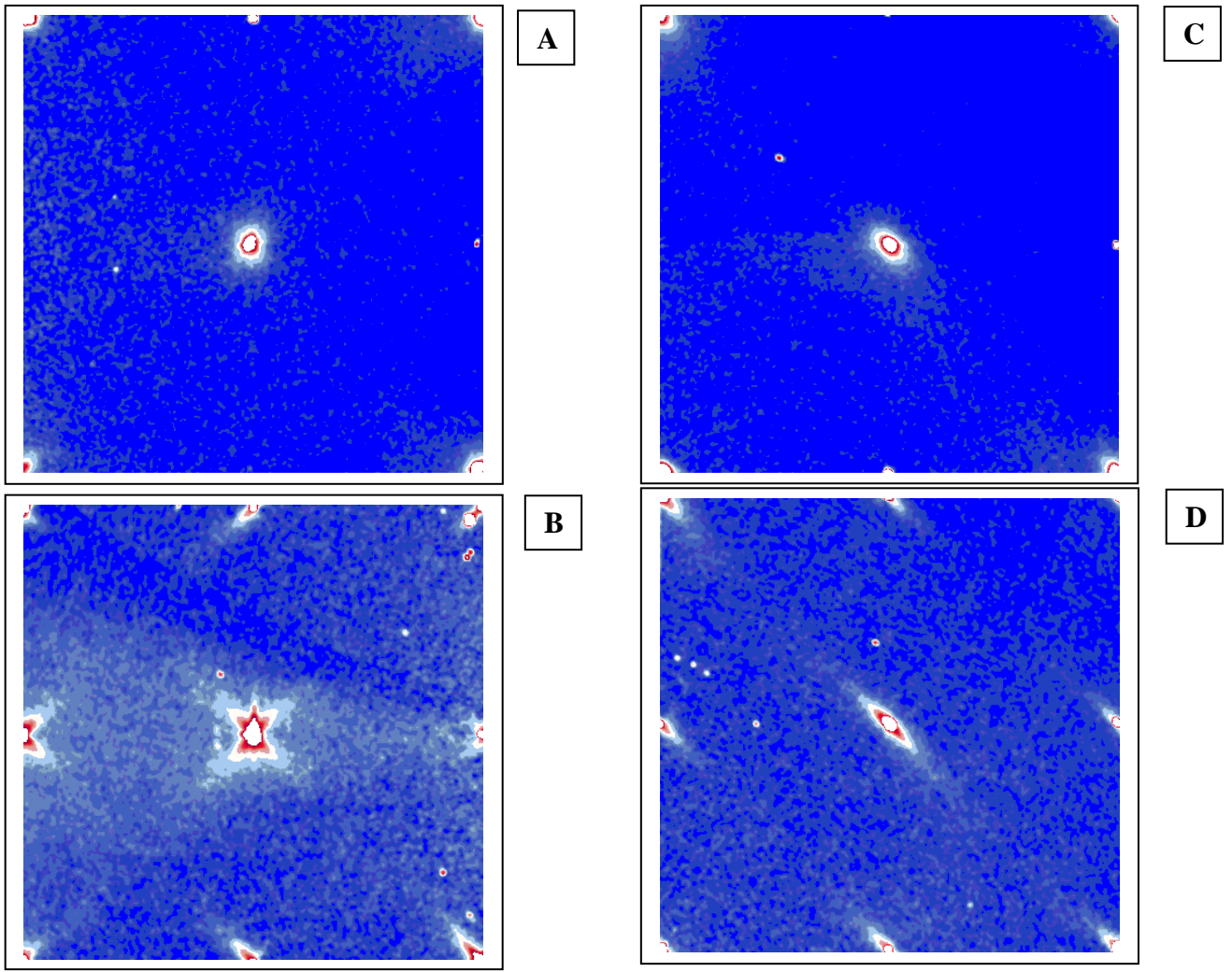

Figure 4. (Color online) Vicinity of the Bragg peaks (2, 0, 0) (left column) and (2, 2, 0) (right column) taken from BMT at 300 K (A.C) and from PMT at 175 K (B.D) [5]. The data was taken at SNBL. The corners of each subplot are the nearest Bragg peak. Taking into account that the lattice parameters of PMT (4.04 Å) and BMT (4.08 Å) are very close one may compare directly the extent and the shapes of the scattering in the two crystals. However, the intensities should not be compared as the appropriate corrections and normalizations were not performed.

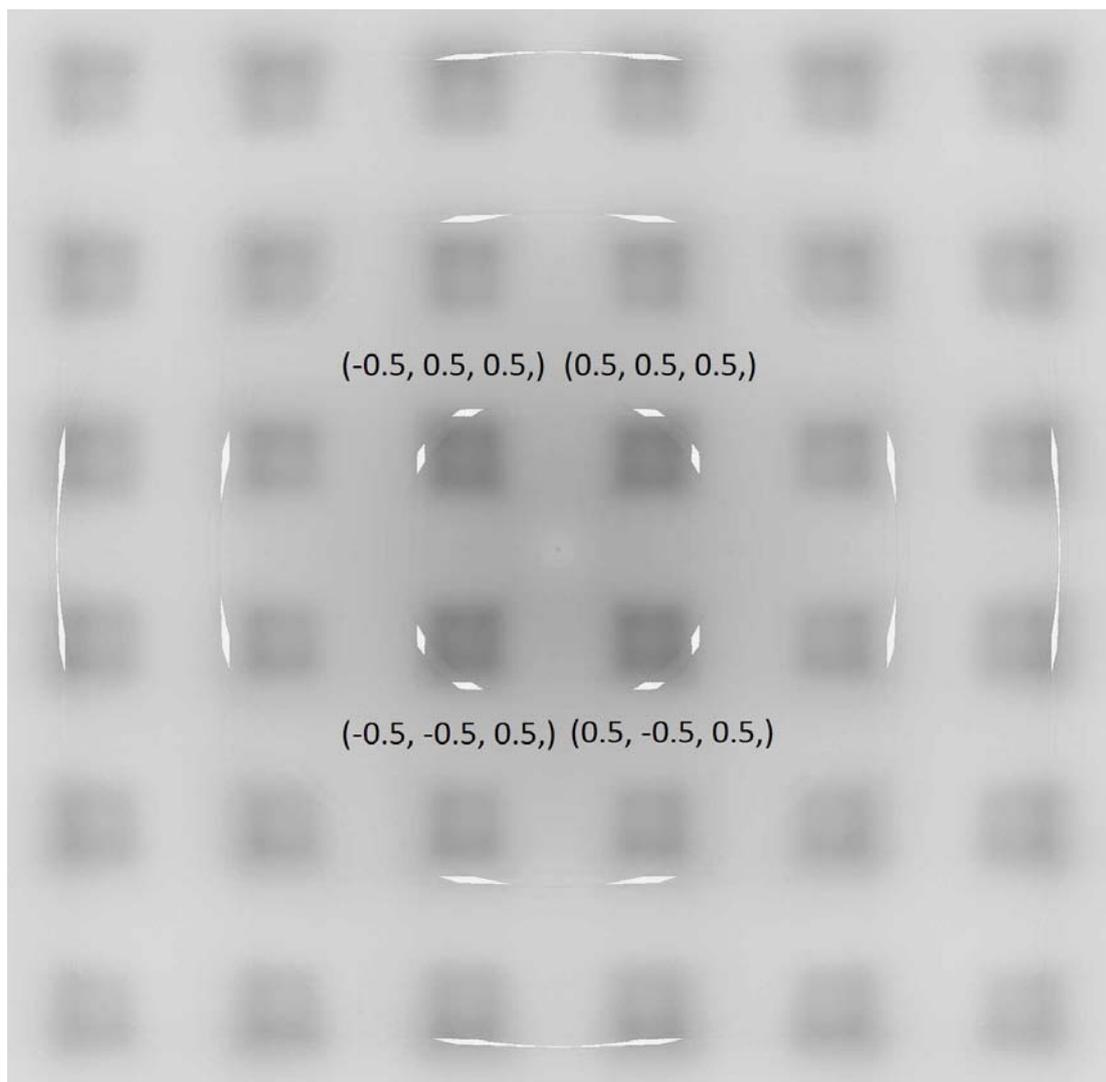

Figure 5. (Color online) Diffuse scattering patterns observed for BaMg$_{1/3}$Ta$_{2/3}$O$_3$ at T = 100 K. The plane is a section through the (H, K, 0.5) plane; the data was taken on the X06SA at SLS. The centres of the squares closest to the origin of the reciprocal space are marked so to show the coordinate system. To facilitate the inspection of the diffuse scattering, the Laue symmetry was used for the reconstruction of this particular reciprocal plane from the experimental patterns. The asymmetry of intensities with respect to the origin of the reciprocal space is due to the irregular shape of the sample.

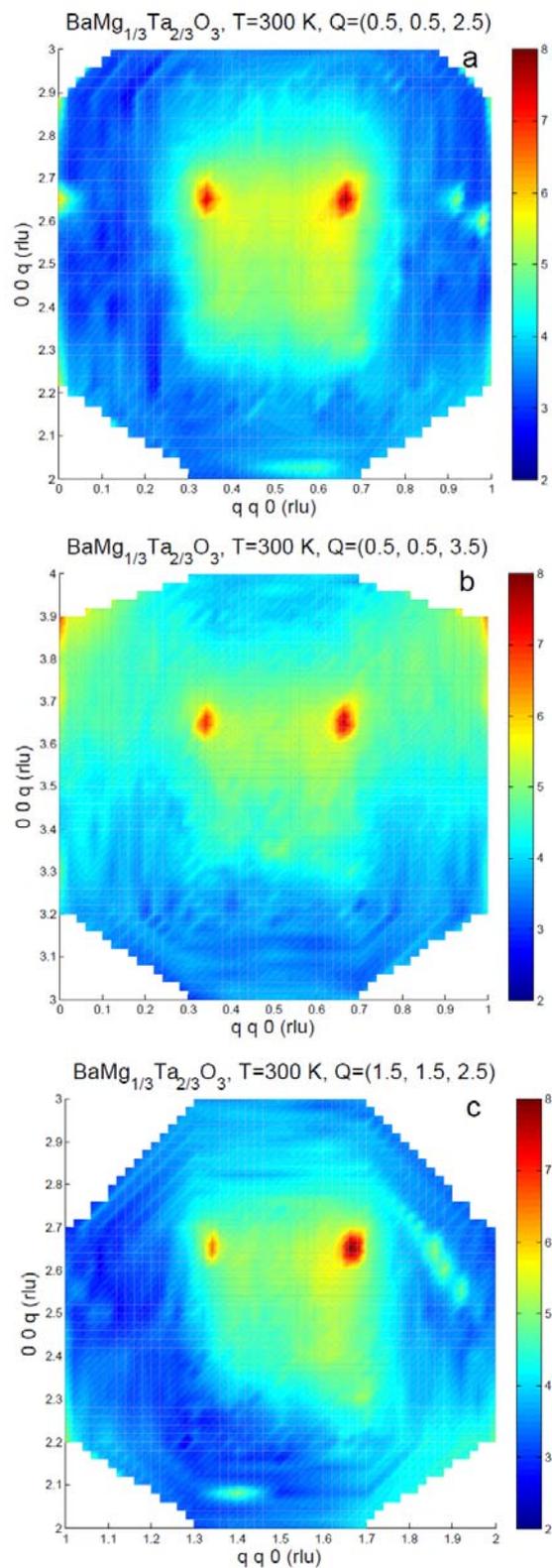

Figure 6. (Color online) Contour plots of the X-ray diffuse scattering from BMT as measured on the XMAS instrument at ESRF. The scattering planes is [H,H,0]/[0,0,L]. Shown Q refers to the central point of each subplot.